\documentclass[12pt,notitlepage,fleqn]{article}
\usepackage{amssymb}

\usepackage{graphicx}
\usepackage{amsmath}
\usepackage{indentfirst}

\input{tcilatex}

\begin{document}

\begin{center}
\ {\Huge \ Deducing the Lowest Rest Mass}\ {\Huge Quarks and Baryons of all
Kinds with an Expanded }

{\Huge Form of Planck-Bohr's Quantization Method}

\bigskip

{\normalsize Jiao Lin Xu}

{\small The Center for Simulational Physics, The Department of Physics and
Astronomy}

{\small University of Georgia, Athens, GA 30602, USA}

E- mail: {\small \ Jxu@Hal.Physast.uga.edu}

\bigskip

\textbf{Abstract}
\end{center}

\bigskip

{\small \ Using an expanded form of Planck-Bohr's quantization method and
phenomenological formulae, we deduce the rest masses and intrinsic quantum
numbers (I, S, C, b and Q) of the lowest energy quarks and baryons of all
kinds, from only one elementary quark family }$\epsilon ${\small \ with
S=C=b=0. The deduced quantum numbers match those found in experiments, and
the deduced rest masses are consistent with experimental results. This paper
predicts some quarks }$\text{ u}_{c}\text{(6073), }${\small d}$_{s}${\small %
(9613) }$\text{and d}_{b}\text{(9333) }${\small and baryons }$\Lambda
_{c}^{+}${\small (6699), }$\Lambda _{b}^{0}${\small (9959) and }$\Lambda $%
{\small (10239).\ }

{\small PACS numbers: 12.39.-x; 14.65.-q; 14.20.-c\ \ \ \ \ \ \ \ \ \ \ \ \
\ \ \ \ \ \ \ \ \ \ \ \ \ \ \ \ \ \ \ \ \ \ \ \ \ \ \ \ \ \ \ \ \ \ \ \ \ \
\ \ \ \ \ \ \ \ \ \ \ \ \ \ \ \ \ \ \ \ \ \ \ \ \ \ \ \ \ \ \ \ \ \ \ \ \ \
\ \ \ \ \ \ \ }

\section{Introduction}

One hundred years ago, most physicists thought that the study of physics had
been \textquotedblleft completed.\textquotedblright\ They believed that it
could explain \textquotedblleft all\textquotedblright\ physical phenomena.
Black body spectrum, however, could not be explained by the physics of that
time, leading Planck to propose a quantization postulate to solve this
problem \cite{Planck}. This quantization eventually led to quantum mechanics.

Today we face a similar situation. \ The standard model \cite{Standard} ``is
in excellent accord with almost all current data.... It has been enormously
successful in predicting a wide range of phenomena,''\ but\ it cannot deduce
the mass spectra of quarks, baryons and mesons. So far, no theory can
successfully do so. This case hints that physics need a ``more fundamental
theory''\ \cite{Standard} than the standard model. What is it? Nobody knows!
Thus we should encourage attempts of all kinds. We try to deduce the spectra
using the expanded form of Planck-Bohr's \cite{Bohr} quantization method and
phenomenological formula. We work with systems (quarks and hadrons) that are
a level deeper than the system (atoms and nuclei) faced by Planck and Bohr.
Therefore, if Planck and Bohr got correct quantizations for atoms and nuclei
using only one simple quantization, we must use two steps and more complex
quantizations. It is worth emphasizing that we must expand upon
Planck-Bohr's\ quantization, performing it twice rather than once, in order
to obtain the short-lived and scarce quarks (a deeper quantization than
atoms and nuclei). Hopefully, this method will help physicists to discover a
more fundamental theory which underlies the standard model \cite{Standard},
just as Planck-Bohr' method has done.

\section{The Elementary Quarks in the Vacuum and Their Free Excited Quarks}

1. We assume that there is only one elementary quark family $\epsilon $ with
two isospin states ($\epsilon _{u}$ has I$_{Z}$ = $\frac{1}{2}$ and Q = +$%
\frac{2}{3}$, $\epsilon _{d}$ has I$_{Z}$ = $\frac{-1}{2}$ and Q = -$\frac{1%
}{3}$). For $\epsilon _{u}$ (or $\epsilon _{d}$),\ there are three colored
quarks. Thus, there are six Fermi (s = $\frac{1}{2}$) elementary quarks in
the $\epsilon $ family with S = C = b = 0 in the vacuum.

2. As a colored elementary quark $\epsilon _{u}$(or $\epsilon _{d}$) is
excited from the vacuum, its color, electric charge and spin do not change,
but it will get energy V (the minimum excited energy from the vacuum).\ The
excited state\ of the elementary quark $\epsilon _{u}$\ is the u-quark with
Q = $\frac{2}{3}$ and s = $\frac{1}{2}$. The excited state\ of the
elementary quark $\epsilon _{d}$\ is the d-quark with Q = - $\frac{1}{3}$
and s = $\frac{1}{2}$. For the excited quark free motion, in general case,
we shall use the Dirac equation. Our purpose is to find the rest masses of
the excited quarks. The rest masses are the low energy limits of the excited
quark masses. If we omit the spin of the quark, the low energy limit of the
Dirac equation is the Schr\"{o}dinger equation. When we use the
Schr\"{o}dinger equation to approach the Dirac equation, we cannot forget
the static energy of the excited quark. We will deal with this energy as a
potential energy (V). The approximate Schr\"{o}dinger equation is:

\begin{equation}
\frac{\hslash ^{2}}{\text{2m}_{\epsilon }}\nabla ^{2}\psi \text{ + (}{\huge %
\varepsilon }\text{-V)}\psi \text{ = 0}  \label{Schrodinger}
\end{equation}
where m$_{\epsilon }$ is the unknown mass of the elementary quark $\epsilon $%
. Omitting electromagnetic mass, we assume m$_{\epsilon }$\TEXTsymbol{>}%
\TEXTsymbol{>}M$_{p}$ = 938 Mev; this is one of the reasons we use the
Schr\"{o}dinger equation instead of the Dirac equation. Our results will
show that this approach is a very good approximation. V is the static
energy, and it is the minimum excited energy of an elementary quark from the
vacuum. The solution of (\ref{Schrodinger}) is the eigen wave function and
the eigen energy of the u-quark or the d-quark:

\begin{equation}
\begin{tabular}{l}
$\psi _{\overrightarrow{k}}\text{(}\overrightarrow{\text{r}}\text{) }%
\backsim \text{ exp(i}\overrightarrow{k}\cdot \overrightarrow{r}\text{),}$
\\ 
$\varepsilon \text{ = V +}\frac{\hslash ^{2}}{2m_{\epsilon }}\text{[(k}_{1}%
\text{)}^{2}\text{+(k}_{2}\text{)}^{2}\text{+(k}_{3}\text{)}^{2}\text{].}$%
\end{tabular}
\label{Wave+Energy}
\end{equation}
According to the Quark Model \cite{Quark Model} a proton p = uud and a
neutron n = udd, omitting electromagnetic mass of quarks, from (\ref
{Wave+Energy}), at $\overrightarrow{k}$ = 0, we have

\begin{eqnarray}
\text{M}_{p} &\text{=}&\text{m}_{u}\text{+m}_{u}\text{+m}_{d}\text{-}%
\left\vert \text{E}_{bind}\right\vert \approx \text{M}_{n}\text{=m}_{u}\text{%
+m}_{d}\text{+m}_{d}\text{-}\left\vert \text{E}_{bind}\right\vert \text{=
939 Mev }  \label{939} \\
&\rightarrow &\text{m}_{u}\text{ = m}_{d}\text{ = V =\ }\frac{1}{3}\text{%
(939 +}\left\vert \text{E}_{bind}\right\vert \text{) = 313 +}\Delta \text{
(Mev)}  \label{313}
\end{eqnarray}
where E$_{bind}$ is the total binding energy of the three quarks in a
baryon. $\Delta $ represents $\frac{1}{3}\left\vert \text{E}%
_{bind}\right\vert $, and is an unknown large positive constant for all
baryons. 
\begin{equation}
\Delta \text{ = }\frac{1}{3}\left\vert \text{E}_{bind}\right\vert \text{ 
\TEXTsymbol{>}\TEXTsymbol{>} M}_{p}\text{.}  \label{Dalta}
\end{equation}
Now that we have free excited the u(313+$\Delta $)-quark and the d(313+$%
\Delta $)-quark, they are long-lived and common quarks; how, then, do we
deduce the short-lived and scarce quarks?\ 

\section{ Deducing Energy Bands With an Expanded Form of Planck-Bohr's
Quantization Method}

1. In order to deduce the short-lived and scarce quarks with the expanded
quantization method, we recall the works of Planck and Bohr.

Planck's \cite{Planck} energy quantization postulate states that
\textquotedblleft any physical entity whose single `coordinate'\ execute
simple harmonic oscillations (i.e., is a sinusoidal function of time) can
possess only total energy $\varepsilon $ which satisfy the relation $%
\varepsilon =nh\nu $, n = 0, 1, 2, 3, .... where $\nu $ is the frequency of
the oscillation and h is a universal constant.\textquotedblright\ Planck
selects reasonable energy from a continuous energy spectrum. \ \ \ \ \ \ \ \
\ \ \ \ 

Bohr's \cite{Bohr} orbit quantization tell us that ``an electron in an atom
moves in a circular orbit about the nucleus...obeying the laws of classical
mechanics, But instead of the infinity of orbits which would be possible in
classical mechanics, it is only possible for an electron to move in an orbit
for which its orbital angular momentum L is an integral multiple of Planck's
constant h, divided by 2$\pi $.''\ Using the quantization condition (L = $%
\frac{nh}{2\pi }$, \ n = 1, 2, 3, ... ), Bohr selects reasonable orbits from
the infinite orbits.

Drawing from these great physicists' works, we find the most important law
is to use quantized conditions and symmetries (as circular orbit) to select
reasonable energy levels from a continuous energy spectrum. \ \ \ \ \ \ \ \
\ \ \ \ 

2. In order to get the short-lived scarce quarks, we quantize the free
motion of an excited quark (\ref{Wave+Energy}) to select energy bands from
the continuous energy.\ The energy bands correspond to short-lived and
scarce quarks.

a. For free motion of an excited quark with continuous energy (\ref
{Wave+Energy}),\ we assume the wave vector $\overrightarrow{k}$ has the
symmetries of the regular rhombic dodecahedron in $\overrightarrow{k}$-space
(see Fig. 1). We assume that the axis $\Gamma $-H in Fig.1 has length $\frac{%
2\pi }{a}$, with an unknown constant a.

\begin{figure}[h]
\vspace{5.8in} \includegraphics{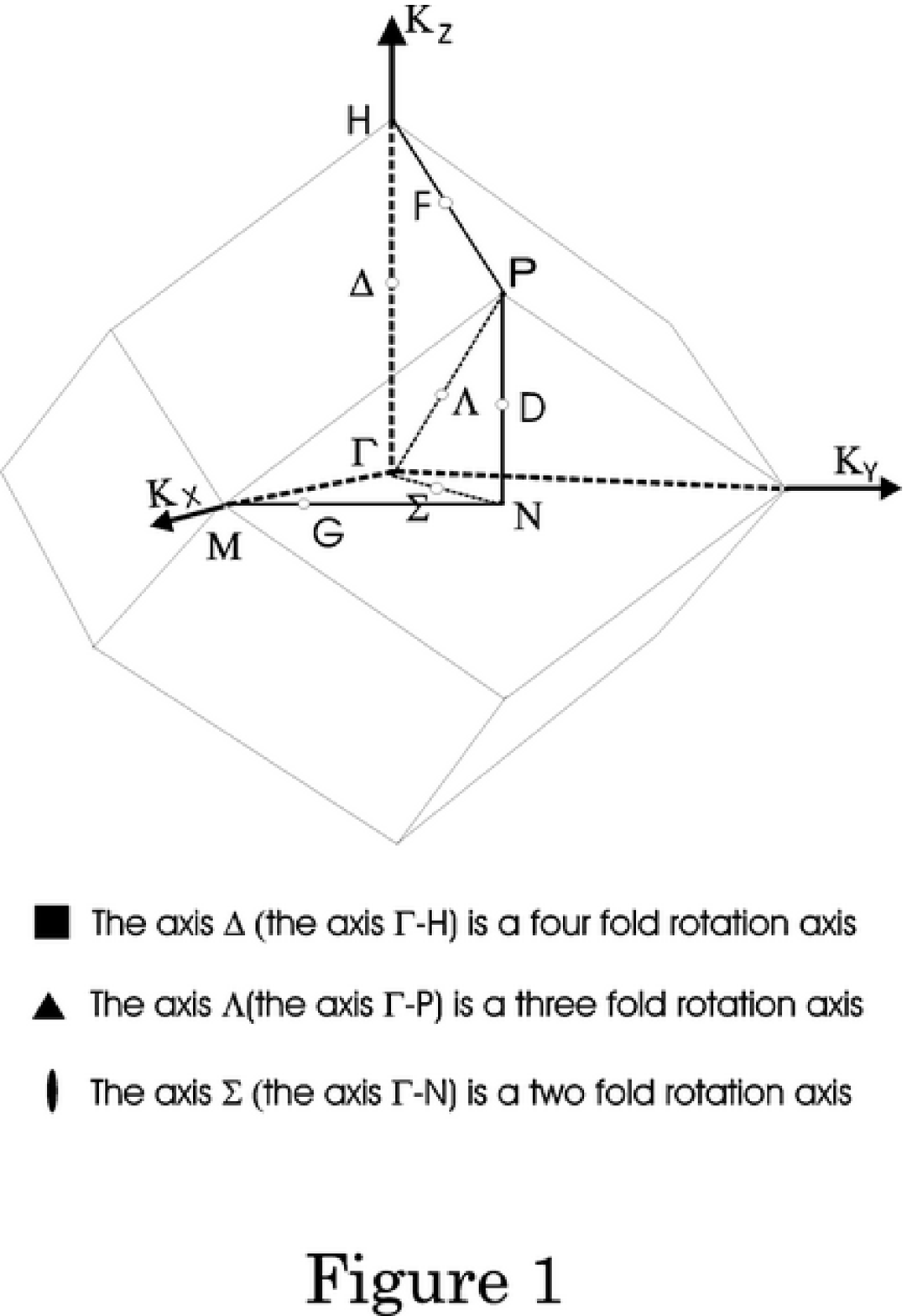} \label{Fig1}
\caption{{\protect\small The regular rhombic dodecahedron. The symmetry
points and axes are indicated. The }$\Delta ${\protect\small -axis is a
fourfold rotation axis, the strange number S = 0 and the fourfold baryon
family }$\Delta ${\protect\small \ (}$\Delta ^{++},${\protect\small \ }$%
\Delta ^{+},${\protect\small \ }$\Delta ^{0},${\protect\small \ }$\Delta
^{-} ${\protect\small ) will appear on the axis. The axes }$\Lambda $%
{\protect\small \ and F are threefold rotation axes, the strange number S =\
-1 and the threefold baryon family }$\Sigma ${\protect\small \ (}$\Sigma
^{+},${\protect\small \ }$\Sigma ^{0},${\protect\small \ }$\Sigma ^{-})$%
{\protect\small \ will appear on the axes. The axes }$\Sigma $%
{\protect\small \ and G are twofold rotation axes, the strange number S = -2
and the twofold baryon family }$\Xi ${\protect\small \ (}$\Xi ^{0},$%
{\protect\small \ }$\Xi ^{-}${\protect\small ) will appear on the axes. The
axis D is parallel to the axis }$\Delta ${\protect\small , S = 0. And it is
a twofold rotation axis, the twofold baryon family N (N}$^{+}$%
{\protect\small , N}$^{0}${\protect\small ) will be on the axis.}}
\end{figure}
The expanded quantizing conditions are: 
\begin{equation}
\begin{tabular}{l}
k$_{1}$ = $\frac{2\pi }{a}$(n$_{1}$-$\xi $), \\ 
k$_{2}$ = $\frac{2\pi }{a}$(n$_{2}$-$\eta $), \\ 
k$_{3}$ = $\frac{2\pi }{a}$(n$_{3}$-$\zeta $).
\end{tabular}
\label{K(1,2,3)}
\end{equation}
Putting (\ref{K(1,2,3)}) into (\ref{Wave+Energy}), we get (\ref{BandW})$\ $%
and (\ref{E(nk)}):

\begin{eqnarray}
\psi _{\overrightarrow{k}}\text{(}\overrightarrow{\text{r}}\text{) }
&\backsim &\text{ exp}\frac{2\pi }{a}\text{[(n}_{1}\text{-}\xi \text{)x + (n}%
_{2}\text{ - }\eta \text{)y+(n}_{3}\text{ - }\zeta \text{)z]}  \label{BandW}
\\
\text{E(}\vec{k}\text{,}\vec{n}\text{) } &\text{=}&\text{ 313 + }\Delta 
\text{ + }\alpha \text{[(n}_{1}\text{-}\xi \text{)}^{2}\text{+(n}_{2}\text{-}%
\eta \text{)}^{2}\text{+(n}_{3}\text{-}\zeta \text{)}^{2}\text{].}
\label{E(nk)}
\end{eqnarray}
Where $\alpha $ =$\frac{\text{h}^{2}}{\text{2m}_{\epsilon }\text{a}^{2}}$.
For $\overrightarrow{n}$ = (n$_{1}$, n$_{2}$, n$_{3}$), n$_{1}$, n$_{2}$ and
n$_{3}$ are $\pm \func{integer}$s and zero. The $\overrightarrow{\kappa }$ =
($\xi $, $\eta $, $\zeta $) has the symmetries of a regular rhombic
dodecahedron. In order to deduce the deeper short-lived and scarce quarks,
we must further quantize the $\overrightarrow{n}$ as follows.

b. Where n$_{1}$, n$_{2}$ and n$_{3}$ are integers to satisfy the expanded
quantizing condition (\ref{l-n}): If we assume n$_{1}$ = \textit{l}$_{2}$ 
\textit{+ l}$_{3}$, n$_{2}$ =\textit{\ l}$_{3}$ \textit{+ l}$_{1}$ and n$%
_{3} $ =\textit{\ l}$_{1}$ \textit{+ l}$_{2},$ so that 
\begin{equation}
\begin{tabular}{l}
\textit{l}$_{1}$ = $\frac{1}{2}$(-n$_{1}$ + n$_{2}$ + n$_{3}$) \\ 
\textit{l}$_{2}$ = $\frac{1}{2}$(+n$_{1}$ - n$_{2}$ + n$_{3}$) \\ 
\textit{l}$_{3}$ = $\frac{1}{2}$(+n$_{1}$ + n$_{2}$ - n$_{3}$),
\end{tabular}
\label{l-n}
\end{equation}
the condition is that only those values of $\overrightarrow{n}$ = (n$_{1}$, n%
$_{2}$, n$_{3}$) are allowed that make $\overrightarrow{l}$ = \textit{(l}$%
_{1}$\textit{, l}$_{2}$\textit{, l}$_{3}$\ ) an integer vector. This is the
expanded quantization for $\vec{n}$ \ values. For example, $\vec{n}$ \
cannot take the values (1, 0, 0) or (1, 1, -1), but can take (0, 0, 2) and
(1, -1, 2). The low level allowed n = (n$_{1}$, n$_{2}$, n$_{3})$ values are
shown in the following table:

\ 

$
\begin{tabular}{|l|}
\hline
{\small E(}$\overrightarrow{n}${\small ,0) = 0\ : (0, 0, 0) \ \ \ \ \ \ \ \
\ \ Notes: }[$\overline{\text{{\small 1}}}${\small 12 }$\equiv $ (-1,1,2)
and $\overline{\text{{\small 1}}}${\small 1}$\overline{\text{{\small 2}}}%
\equiv $ (-1,1,-2)] \\ \hline
{\small E(}$\overrightarrow{n}${\small ,0) = 2\ :}$\ ${\small (101, }$%
\overline{\text{{\small 1}}}${\small 01, 011, 0}$\overline{\text{{\small 1}}}
${\small 1, 110, 1}$\overline{\text{{\small 1}}}${\small 0, }$\overline{%
\text{{\small 1}}}${\small 10, }$\overline{\text{{\small 1}}}\overline{\text{%
{\small 1}}}${\small 0, 10}$\overline{\text{{\small 1}}}${\small , }$%
\overline{\text{{\small 1}}}${\small 0}$\overline{\text{{\small 1}}}${\small %
, 01}$\overline{\text{{\small 1}}}${\small , 0}$\overline{\text{{\small 1}}}%
\overline{\text{{\small 1}}}${\small )} \\ \hline
{\small E(}$\overrightarrow{n}${\small ,0) = 4\ :\ \ (002, 200}, {\small 200}%
$\text{, }\overline{\text{{\small 2}}}\text{{\small 00}, }${\small 0}$%
\overline{\text{{\small 2}}}${\small 0, 00}$\overline{\text{{\small 2}}}$%
{\small )} \\ \hline
{\small E(}$\overrightarrow{n}${\small ,0) = 6:} 
\begin{tabular}{l}
{\small 112, 211, 121, }$\overline{\text{{\small 1}}}${\small 21,}$\overline{%
\text{{\small 1}}}${\small 12, 2}$\overline{\text{{\small 1}}}${\small 1}, 
{\small 1}$\overline{\text{{\small 1}}}${\small 2, 21}$\overline{\text{%
{\small 1}}}${\small ,12}$\overline{\text{{\small 1}}}${\small ,}$\overline{%
\text{{\small 2}}}${\small 11, 1}$\overline{\text{{\small 2}}}${\small 1, 11}%
$\overline{\text{{\small 2}}}$,{\small \ } \\ 
$\overline{\text{{\small 11}}}${\small 2, }$\overline{\text{{\small 1}}}$%
{\small 2}$\overline{\text{{\small 1}}}$,{\small \ 2}$\overline{\text{%
{\small 11}}}${\small , }$\overline{\text{{\small 21}}}${\small 1}, $%
\overline{\text{{\small 12}}}${\small 1, 1}$\overline{\text{{\small 12}}}$%
{\small , 1}$\overline{\text{{\small 21}}}${\small ,}$\overline{\text{%
{\small 1}}}${\small 1}$\overline{\text{{\small 2}}}${\small ,}$\overline{%
\text{{\small 2}}}$1$\overline{\text{{\small 1}}}$,{\small \ }$\overline{%
\text{{\small 211}}}${\small , }$\overline{\text{{\small 121}}},${\small \ }$%
\overline{\text{{\small 112}}},$%
\end{tabular}
\\ \hline
\end{tabular}
$

\ \ \ \ \ \ \ \ \ \ \ \ \ 

c. The\ vector $\overrightarrow{\kappa }$ = ($\xi $, $\eta $, $\zeta $) in (%
\ref{BandW}) and\ (\ref{E(nk)}) has the symmetries of the regular rhombic
dodecahedron in k-space (see Fig. 1). From Fig. 1, we can see that there are
four kinds of symmetry points ($\Gamma $, H, P and N) and six kinds of
symmetry axes ($\Delta $, $\Lambda $, $\Sigma $, D, F and G) in the regular
rhombic dodecahedron. The coordinates ($\xi $, $\eta $, $\varsigma $) of the
symmetry axes are: 
\begin{equation}
\begin{tabular}{ll}
$\Delta \text{ }\text{= (0, 0, }\zeta \text{),\ 0 \TEXTsymbol{<}}\zeta \text{
\TEXTsymbol{<}1; }$ & $\Lambda \text{ = (}\xi \text{, }\xi \text{, }\xi 
\text{), 0 \TEXTsymbol{<}}\xi \text{ \TEXTsymbol{<}}\frac{\text{1}}{\text{2}}%
\text{;}$ \\ 
$\Sigma \text{{} }\text{= (}\xi \text{, }\xi \text{, 0), 0 \TEXTsymbol{<}}%
\xi \text{ \TEXTsymbol{<}}\frac{\text{1}}{\text{2}}\text{;}$ & $\text{D }%
\text{= (}\frac{\text{1}}{\text{2}}\text{, }\frac{\text{1}}{\text{2}}\text{, 
}\xi \text{), 0 \TEXTsymbol{<}}\xi \text{ \TEXTsymbol{<}}\frac{\text{1}}{%
\text{2}}\text{;}$ \\ 
$\text{G }\text{= (}\xi \text{, 1-}\xi \text{, 0), }\frac{\text{1}}{\text{2}}%
\text{ \TEXTsymbol{<}}\xi \text{ \TEXTsymbol{<}1;}$ & $\text{F = (}\xi \text{%
, }\xi \text{, 1-}\xi \text{), 0 \TEXTsymbol{<}}\xi \text{ \TEXTsymbol{<}}%
\frac{\text{1}}{\text{2}}\text{.}$%
\end{tabular}
\label{Sym-Axes}
\end{equation}
\qquad The energy (\ref{E(nk)}) of the excited quarks has six kinds of
symmetry axes (\ref{Sym-Axes}) (see Table A2 and Table A3 of \cite{0502091}).

d. The energy (\ref{E(nk)}) with a $\overrightarrow{n}$\ = (n$_{1}$, n$_{2}$%
, n$_{3}$) along a symmetry axis (coordinates ($\xi $, $\eta $, $\varsigma $%
) of (\ref{Sym-Axes})) forms an energy band.

e. Each energy band corresponds to a short-lived and scarce quark.\textbf{\ }%
Any excited elementary quark that is not along a symmetry axis\ is the
long-lived u-quark (or the d- quark).

3. After getting (\ref{E(nk)}), (\ref{l-n}) and (\ref{Sym-Axes}), using (\ref
{360}) we can deduce low energy bands of the six symmetry axes (see Table
B1-B3 and Table B4-B7 of \cite{0502091}). As an example, we can show the
single energy bands of the $\Delta $-axis and the $\Sigma $-axis in Table 1:

\begin{tabular}{l}
\ \ \ \ \ \ \ \ \ \ \ \ \ \ \ \ \ \ \ \ \ \ \ \ \ \ \ \ \ \ \ \ \ \ \ \
Table 1. \ The Single Energy Bands \\ 
\begin{tabular}{|l|l|}
\hline
\ \ \ \ \ \ \ \ \ \ \ \ \ \ \ \ \ \ \ The $\Delta $-Axis & \ \ \ \ \ \ \ \ \
\ \ \ \ \ \ \ The $\Sigma $-Axis \\ \hline
$
\begin{tabular}{|l|l|l|l|}
\hline
$\text{E}_{Start}$ & $\text{(n}_{1}\text{,n}_{2}\text{,n}_{3}${\small )} & 
{\small E(}$\vec{k}${\small ,}$\vec{n}${\small )} & E$_{end}$ \\ \hline
$\text{E}_{\Gamma }\text{=0}$ & {\small (0, 0, 0): } & 313 & $\text{E}_{H}%
\text{=1}$ \\ \hline
$\text{E}_{H}\text{=1}$ & {\small (0, 0, 2); }$\text{ }$ & {\small 673} & $%
\text{E}_{\Gamma }\text{=4}$ \\ \hline
$\text{E}_{\Gamma }\text{=4}$ & {\small (0, 0, }$\overline{2}${\small ); } & 
{\small 1753} & $\text{E}_{H}\text{=9}$ \\ \hline
$\text{E}_{H}\text{=9}$ & {\small (0, 0, 4); }$\text{ }$ & {\small 3553} & $%
\text{E}_{\Gamma }\text{=16}$ \\ \hline
$\text{E}_{\Gamma }\text{=16}$ & {\small (0, 0, }$\overline{\text{{\small 4}}%
}${\small ); } & {\small 6073} & $\text{E}_{H}\text{=25}$ \\ \hline
$\text{E}_{H}\text{=25}$ & (0, 0, 6) & {\small 9313} & $\text{E}_{\Gamma }%
\text{=36}$ \\ \hline
&  &  &  \\ \hline
&  &  &  \\ \hline
\end{tabular}
$ & $
\begin{tabular}{|l|l|l|l|}
\hline
{\small E}$_{Point}$ & $\text{\ n}_{1}\text{n}_{2}\text{n}_{3}$ & {\small E(}%
$\overrightarrow{k}${\small ,}$\overrightarrow{n}${\small )} & E$_{end}$ \\ 
\hline
$\text{E}_{\Gamma }\text{=0}$ & {\small (0, 0, 0)} & {\small 313} & $\text{E}%
_{N}\text{=}\frac{\text{1}}{2}$ \\ \hline
$\text{E}_{N}\text{=}\frac{\text{1}}{2}$ & $(\text{1,1,0})$ & $\text{493}$ & 
$\text{E}_{\Gamma }\text{=2}$ \\ \hline
$\text{E}_{\Gamma }\text{=2}$ & $(\text{-1,-1,0})$ & $\text{1033}$ & $\text{E%
}_{N}\text{=}\frac{\text{9}}{2}$ \\ \hline
$\text{E}_{N}\text{=}\frac{\text{9}}{2}$ & $(\text{2,2,0})$ & $\text{1933}$
& $\text{E}_{\Gamma }\text{=8}$ \\ \hline
$\text{E}_{\Gamma }\text{=8}$ & $(\text{-2,-2,0})$ & $\text{3193}$ & $\text{E%
}_{N}\text{=}\frac{\text{25}}{2}$ \\ \hline
$\text{E}_{N}\text{=}\frac{\text{25}}{2}$ & $(\text{3,3,0}) $ & $\text{4813}$
& $\text{E}_{\Gamma }\text{=18}$ \\ \hline
$\text{E}_{\Gamma }\text{=18}$ & {\small (-}3,-3,0{\small )} & $\text{6793}$
& $\text{E}_{N}\text{=}\frac{\text{49}}{2}$ \\ \hline
$\text{E}_{N}\text{=}\frac{\text{49}}{2}$ & (4, 4, 4) & $\text{9133}$ & $%
\text{E}_{\Gamma }\text{=32}$ \\ \hline
\end{tabular}
$ \\ \hline
\end{tabular}
\end{tabular}
\ \ \ \ \ \ \ \ \ \ \ \ \ \ \ \ \ \ \ \ \ \ \ \ \ \ \ \ \ \ 

\section{Deducing the Lowest Energy Quarks of all Kinds\ with
Phenomenological Formulae\ \ \ \ \ \ \ \ \ \ \ \ \ \ \ \ \ \ \ \ \ \ \ \ \ \
\ \ \ \ \ \ \ \ \ \ \ \ \ \ \ \ \ \ \ \ \ \ \ \ \ \ \ \ \ \ \ \ \ \ \ \
\qquad \qquad \qquad \qquad\ \ \ \ \ \ \ \ \ \ \ \ \ \ \ \ \ \ \ \ \ \ \ \ \
\ \ \ \ \ \ \ \ \ \ \ \ \ \ \ \ \ \ \ \ \ \ \ \ \ \ \ \ \ \ \ \ \ \ \ \ \ \
\ \ \ \ \ \ \ \ \ \ \ \ \ \ \ \ \ \ \ \ \ \ \ \ \ \ \ \ \ \ \ \ \ \ \ \ \ \
\ \ \ \ \ \ \ \ \ \ \ \ \ \ \ \ \ \ \ \ \ \ \ \ \ \ \ \ \ \ \ \ \ \ \ \ \ \
\ \ \ \ \ \ \ \ \ \ \ \ \ \ \ \ \ \ \ \ \ \ \ \ \ \ \ \ \ \ \ \ \ \ \ \ \ \
\ \ \ \ \ \ \ \ \ \ \ \ \ \ \ \ \ \ \ \ \ \ \ \ \ \ \ \ \ \ \ \ \ \ \ \ \ \
\ \ \ \ \ \ \ \ \ \ \ \ \ \ \ \ \ \ \ \ \ \ \ \ \ \ \ \ \ \ \ \ \ \ \ \
\qquad\ \ \ \ \ \ \ \ \ \ \ \ \ \ \ \ \ \ \ \ \ \ \ \ \ \ \ \ \ \ \ \ \ \ \
\ \ \ \ \ \ \ \ \ \ \ \ \ \ \ \ \ \ \ \ \ \ \ \ \ \ \ \ \ \ \ \ \ \ \ \ \ \
\ \ \ \ \ \ \ \ \ \ \ \ \ \ \ \ \ \ \ \ \ \ \ \ \ \ \ \ \ \ \ \ \ \ \ \ \ \
\ \ \ \ \ \ \ \ \ \ \ \ \ \ \ \ \ \ \ \ \ \ \ \ \ \ \ \ \ \ \ \ \ \ \ \ \ \
\ \ \ \ \ \ \ \ \ \ \ \ \ \ \ \ \ \ \ \ \ \ \ \ \ \ \ \ \ \ \ \ \ \ \ \ \ \
\ \ \ \ \ \ \ \ \ \ \ \ \ \ \ \ \ \ \ \ \ \ \ \ \ \ \ \ \ \ \ \ \ \ \ \ \ \
\ \ \ \ \ \ \ \ \ \ \ \ \ \ \ \ \ \ \ \ \ \ \ \ \ \ \ \ \ \ \ \ \ \ \ \ \ \
\ \ \ \ \ \ \ \ \ \ \ \ \ \ \ \ \ \ \ \ \ \ \ \ \ \ \ \ \ \ \ \ \ \ \ \ \ \
\ \ \ \ \ \ \qquad\ \ \ \ \ \ \ \ \ \ \ \ \ \ \ \ \ \ \ \ \ \ \ \ \ \ \ \ \
\ \ \ \ \ \ \ \ \ \ \ \ \ \ \ \ \ \ \ \ \ \ \ \ \ \ \ \ \ \ \ \ \ \ \ \ \ }

\subsection{The Phenomenological Formulae for Intrinsic Quantum Numbers\ of
Quarks\ \ \ \ \ \ \ \ \ \ \ \ \ \ \ \ \ \ \ \ \ \ \ \ \ \ \ \ \ \ \ \ \ \ \
\ \ \ \ \ \ \ \ \ \ \ \ \ \ \ \ \ \ \ \ \ \ \ \ \ \ \ \ \ \ \ \ \ \ \ \ \ \
\ \ \ \ \ \ \ \ \ \ \ \ \ \ \ \ \ \ \ \ \ \ \ \ \ \ \ \ \ \ \ \ \ \ \ \ \ \
\ \ \ \ \ \ \ \ \ \ \ \ \ \ \ \ \ \ \ \ \ \ \ \ \ \ \ \ \ \ \ \ \ \ \ \ \ \
\ \ \ \ \ \ \ \ \ \ \ \ \ \ \ \ \ \ \ \ \ \ \ \ \ \ \ \ \ \ \ \ \ \ \ \ \ \
\ \ \ \ \ \ \ \ \ \ \ \ \ \ \ \ \ \ \ \ \ \ \ \ \ \ \ \ \ \ \ \ \ \ \ \ \ \
\ \ \ \ \ \ \ \ \ \ \ \ \ \ \ \ \ \ \ \ \ \ \ \ \ \ \ \ \ \ \ \ \ \ \ \ \ \
\ \ \ \ \ \ \ \ \ \ \ \ \ \ \ \ \ \ \ \ \ \ \ \ \ \ \ \ \qquad\ \ \ \ \ \ \
\ \ \ \ \ \ \ \ \ \ \ \ \ \ \ \ \ \ \ \ \ \ \ \ \ \ \ \ \ \ \ \ \ \ \ \ \ \
\ \ \ \ \ \ \ \ \ \ \ \ \ \ \ \ \ \ \ \ \ \qquad}

In order to deduce the\textbf{\ }short-lived and scarce quarks\textbf{\ }%
with phenomenological formulae we assume the following phenomenological
formulae:

1). For a group of degenerate energy bands (number=deg) with the same energy
and equivalent $\overrightarrow{n}$ values (66) of \cite{0502091}, the
isospin is

\begin{equation}
\text{I = }\frac{\text{deg - 1}}{\text{2}}  \label{IsoSpin}
\end{equation}

2). The strange number S of an excited quark that lies on an axis with a
rotary fold R of the regular rhombic dodecahedron is 
\begin{equation}
\text{S = R - 4.}  \label{S-Number}
\end{equation}

3). For the single energy bands on the $\Gamma $-H and the $\Gamma $-N axis,
the strange number is 
\begin{equation}
\text{S = S}_{axis}\text{+}\Delta \text{S, \ }\Delta \text{S = }\delta \text{%
(}\widetilde{n}\text{) + [1-2}\delta \text{(S}_{axis}\text{)]Sign(}%
\widetilde{n}\text{)}  \label{S+DS}
\end{equation}
where $\delta $($\widetilde{n}$)\ and $\delta $(S$_{axis}$) are Dirac
functions, and S$_{axis}$ is the strange number (\ref{S-Number}) of the axis
(see Table A3 of \cite{0502091}). For an energy band with $\overrightarrow{n}
$ = (n$_{1}$, n$_{2}$, n$_{3})$, $\widetilde{n}$ \ is defined as 
\begin{equation}
\widetilde{n}\text{ }\equiv \frac{\text{n}_{1}\text{+n}_{2}\text{+n}_{3}}{%
\left\vert \text{n}_{1}\right\vert \text{+}\left\vert \text{n}%
_{2}\right\vert \text{+}\left\vert \text{n}_{3}\right\vert }\text{. Sgn(}%
\widetilde{n}\text{) = }\left[ \text{%
\begin{tabular}{l}
+1 for $\widetilde{n}$ \TEXTsymbol{>} 0 \\ 
0 \ \ for $\widetilde{n}$ = 0 \\ 
-1 \ for $\widetilde{n}$ \TEXTsymbol{<} 0
\end{tabular}
}\right]  \label{n/n}
\end{equation}

\begin{equation}
\text{If }\widetilde{n}\text{ = 0 \ \ \ \ }\Delta \text{S = }\delta \text{%
(0) = +1 \ from (\ref{S+DS}) and (\ref{n/n}).}  \label{n=0-DS=+1}
\end{equation}

\begin{equation}
\text{If \ }\widetilde{n}\text{ = }\frac{0}{0}\text{, }\Delta \text{S = - S}%
_{Axis}\text{ .}  \label{DaltaS=-Sax}
\end{equation}
Thus, for $\overrightarrow{n}$ = (0, 0, 0), from (\ref{DaltaS=-Sax}), we
have \ 
\begin{equation}
\text{S = S}_{Axis}\text{+}\Delta \text{S = S}_{Axis}\text{- S}_{Axis}\text{
= 0.}  \label{S=0 of n=0}
\end{equation}
\qquad \qquad \qquad\ \ \ \ 

4). If S = +1, we call it the charmed number C (= 1): 
\begin{equation}
\text{if }\Delta \text{S = +1}\rightarrow \text{S =S}_{Ax}\text{+}\Delta 
\text{S = +1, }C\text{ }\equiv \text{+1.}  \label{Charmed}
\end{equation}
If S = -1$,$ which originates from $\Delta S=+1$ on a single energy band (S$%
_{Ax}$= -2), and there is an energy fluctuation,$\ $we call it the bottom
number\ b:$\ \ \ \ \ \ \ \ \ \ \ \ \ $%
\begin{equation}
\text{for single bands, if\ }\Delta \text{S = +1}\rightarrow \text{S = -1
and }\Delta \text{E}\neq \text{0, b}\equiv \text{-1.}  \label{Battom}
\end{equation}

5). The sixfold energy bands of the F-axis (R = 3 and S = -1) need two
divisions. In the first division, the sixfold band divides into two ($\frac{6%
}{R}$=2) threefold bands; the energy and the strange number do not change (S
= -1 steel). In the second division (K = 1), the threefold bands with
non-equivalent n values divide into a twofold band and a $\sin $gle band.
For the twofold energy bands with two equivalent $\overline{n}$ values (see
(66) of \cite{0502091}) 
\begin{equation}
\text{\lbrack (for sixfold bands) }\Delta \text{S = +1 and \ E \TEXTsymbol{>}
m}_{u_{c}}\text{(1753 Mev)]}\rightarrow \text{q}_{_{\Xi _{C}}}\text{,}
\label{Kersa-C}
\end{equation}
the twofold band represents a twofold family q$_{\Xi _{C}}$-quark with I = $%
\frac{1}{2}$, S = -1 and C = +1.

The sixfold energy bands of the G-axis (R = 2 and S = -2) also need two
divisions. In the first division, the sixfold band divides into three ($%
\frac{6}{R}$=3) twofold bands. The energy and the strange number do not
change (S = -2). In the second division (K = 1), the twofold band with
non-equivalent $\overrightarrow{n}$ values divides into two single bands.
For a single energy band, 
\begin{equation}
\text{\lbrack (for sixfold bands) if }\Delta \text{S = +1 and \ E 
\TEXTsymbol{>} m}_{u_{c}}\text{(1753 Mev)]}\rightarrow \text{d}_{\Omega _{C}}%
\text{,}  \label{Omiga-C}
\end{equation}
the single energy band represents a d$_{\Omega _{C}}$-quark with I = 0, S =
-2 and C = +1.

6). The elementary quark $\epsilon _{u}$ (or $\epsilon _{d}$) determines the
electric charge Q of an excited quark. For an excited quark of $\epsilon
_{u} $ (or $\epsilon _{d}$), Q = +$\frac{2}{3}$ (or -$\frac{1}{3}$). For an
excited quark with isospin I, there are 2I +1 members . I$_{z}$ \TEXTsymbol{>%
} 0, Q = +$\frac{2}{3}$; I$_{z}$ \TEXTsymbol{<} 0, Q = -$\frac{2}{3}$; 
\begin{eqnarray}
\text{I}_{z}\text{ } &\text{=}&\text{ 0, if S+C+b \TEXTsymbol{>} 0, Q = Q}%
_{\epsilon _{u}(0)}\text{ = }\frac{2}{3}\text{;}  \label{2/3} \\
\text{I}_{z}\text{ } &\text{=}&\text{ 0, if S+C+b \TEXTsymbol{<} 0, Q = Q}%
_{\epsilon _{d}(0)}\text{ = -}\frac{1}{3}\text{.}  \label{- 1/3}
\end{eqnarray}
There is no quark with I$_{z}$ = 0 and S + C + b = 0.\ \ 

7). We assume a fluctuation $\Delta $E of a quark energy is

\begin{equation}
\Delta \text{E = 100 S[(1+S}_{Ax}\text{)(J}_{S,}\text{+S}_{Ax}\text{)]}%
\Delta \text{S \ \ \ J}_{S}\text{=}\left\vert \text{S}_{Ax}\right\vert +%
\text{1,2,3, ....}  \label{Dalta-E}
\end{equation}
Fitting experimental results, we can get 
\begin{equation}
\alpha \text{ = 360 Mev.}  \label{360}
\end{equation}
The rest mass (m$^{\ast }$) of a quark, from (\ref{E(nk)}), (\ref{313}), (%
\ref{360}) and (\ref{Dalta-E}) is 
\begin{equation}
\begin{tabular}{l}
$\text{m}^{\ast }\text{ = \{313+ 360 minimum[(n}_{1}\text{-}\xi \text{)}^{2}%
\text{+(n}_{2}\text{-}\eta \text{)}^{2}\text{+(n}_{3}\text{-}\zeta \text{)}%
^{2}\text{]+}\Delta \text{E+}\Delta \text{\} (Mev)}$ \\ 
\ \ \ \ = m + $\Delta $ \ (Mev),
\end{tabular}
\label{Rest Mass}
\end{equation}
This formula (\ref{Rest Mass}) is the united quark mass formula.

\subsection{Deducing Quarks with the Phenomenological Formulae from Energy
Bands\ \ \ \ \ \ \ \ \ \ \ \ \ \ \ }

Using the above formulae (\ref{IsoSpin})-(\ref{Rest Mass}) and the deduced
energy bands shown in Table B1-B3 and Table B5-B7 of \cite{0502091}, we can
deduce all low energy quarks that are sufficient to cover all experimental
data (see Table 11 of \cite{0502091}). Since the five ground quarks are all
born on the single energy bands of the $\Delta $-axis and the $\Sigma $%
-axis, we deduce the quarks of these single energy bands as examples using
Table 2 and Table 3.

\begin{tabular}{l}
\ \ \ Table 2. The u$_{C}$(m$^{\ast }$)-quarks and the d$_{S}$(m$^{\ast }$%
)-quarks on the $\Delta $-axis (S$_{\Delta }$=0) \\ 
\begin{tabular}{|l|l|l|l|l|l|l|l|l|l|l|}
\hline
$\text{E}_{Point}$ & E$_{(\overrightarrow{k},\overrightarrow{n})}$ & $\text{n%
}_{1,}\text{n}_{2,}\text{n}_{3}$ & $\Delta \text{S}$ & J & I & S & C & Q & $%
\Delta \text{E}$ & $q_{\text{Name}}(m^{\ast })$ \\ \hline
$\text{E}_{\Gamma }\text{=0}$ & 313 & $\text{{\small 0,\ \ 0, \ 0}}$ & 0 & J$%
\text{ =0}$ & $\frac{1}{2}$ & 0 & 0 & $\frac{2}{3}$ & 0 & $\text{u(313+}%
\Delta \text{)}$ \\ \hline
$\text{E}_{H}\text{=1}$ & 673 & $\text{{\small 0, \ 0, \ 2}}$ & -1 & J$_{%
\text{S,H}}\text{ =1}$ & 0 & -1 & 0 & $\frac{-1}{3}$ & 100 & $\text{d}_{S}%
\text{(773+}\Delta \text{)}$ \\ \hline
$\text{E}_{\Gamma }\text{=4}$ & 1753 & $\text{{\small 0, \ 0, -2}}$ & +1 & J$%
_{\text{C,}\Gamma }\text{=1}$ & 0 & 0 & 1 & $\frac{2}{3}$ & 0 & $\text{u}_{C}%
\text{(1753+}\Delta \text{)}$ \\ \hline
$\text{E}_{H}\text{=9}$ & 3553 & $\text{{\small 0, \ 0, \ 4}}$ & -1 & J$_{%
\text{S,H}}\text{ =2}$ & 0 & -1 & 0 & $\frac{-1}{3}$ & 200 & $\text{d}_{S}%
\text{(3753+}\Delta \text{)}$ \\ \hline
$\text{E}_{\Gamma }\text{=16}$ & 6073 & $\text{{\small 0, \ 0, -4}}$ & +1 & J%
$_{\text{C,}\Gamma }\text{=2}$ & 0 & 0 & 1 & $\frac{2}{3}$ & 0 & $\text{u}%
_{C}\text{(6073+}\Delta \text{)}$ \\ \hline
$\text{E}_{H}\text{=25}$ & $\text{9313}$ & $\text{{\small 0, \ 0, \ 6}}$ & -1
& J$_{\text{S,H}}\text{ =3}$ & 0 & -1 & 0 & $\frac{-1}{3}$ & 300 & $\text{d}%
_{S}\text{(9613+}\Delta \text{)}$ \\ \hline
\end{tabular}
\end{tabular}

\begin{tabular}{l}
\ \ \ \ \ \ Table 3. The $\text{d}_{b}$-quarks, d$_{S}$-quarks and d$%
_{\Omega }$-quarks on the $\Sigma $-axis (S$_{\Sigma }$ = -2) \\ 
$
\begin{tabular}{|l|l|l|l|l|l|l|l|l|l|l|}
\hline
{\small E}$_{Point}$ & {\small E}$_{(\overrightarrow{k},\overrightarrow{n})}$
& $\text{\ }${\small n}$_{1}${\small ,n}$_{2,}${\small ,n}$_{3}$ & $\Delta 
\text{S}$ & {\small S} & {\small b} & {\small Q} & $\ \ \text{J}$ & {\small I%
} & $\Delta \text{E} $ & $\text{q}_{Name}\text{(m}^{\ast }\text{)}$ \\ \hline
$\text{E}_{\Gamma }\text{=0}$ & 313 & {\small (0, 0, 0)} & {\small +2} & 
{\small 0} & {\small 0} & $\frac{-1}{3}$ & {\small J}$_{\text{S,}\Gamma }%
\text{ =0}$ & $\frac{1}{2}$ & {\small 0} & {\small d(313}$\text{+}\Delta $%
{\small )} \\ \hline
$\text{E}_{N}\text{=}\frac{\text{1}}{2}$ & $\text{493}$ & $(${\small 1,1,0}$%
) $ & {\small +1} & {\small -1} & {\small 0} & $\frac{-1}{3}$ & {\small J}$_{%
\text{S,N}}\text{ =1}$ & {\small 0} & {\small 0} & $\text{d}_{S}\text{(}$%
{\small 493+}$\Delta \text{)}$ \\ \hline
$\text{E}_{\Gamma }\text{=2}$ & $\text{1033}$ & $(${\small -1,-1,0}$)$ & 
{\small -1} & {\small -3} & {\small 0} & $\frac{-1}{3}$ & {\small J}$_{\text{%
S,}\Gamma }\text{ =1}$ & {\small 0} & {\small 0} & $\text{d}_{\Omega }\text{(%
}${\small 1033+}$\Delta \text{)}$ \\ \hline
$\text{E}_{N}\text{=}\frac{\text{9}}{2}$ & $\text{1933}$ & $(${\small 2,2,0}$%
)$ & {\small +1} & {\small -1} & {\small 0} & $\frac{-1}{3}$ & {\small J}$_{%
\text{S,N}}\text{ =2}$ & {\small 0} & {\small 0} & $\text{d}_{S}\text{(}$%
{\small 1933+}$\Delta \text{)}$ \\ \hline
$\text{E}_{\Gamma }\text{=8}$ & $\text{3193}$ & $(${\small -2,-2}$\text{,0})$
& {\small -1} & {\small -3} & {\small 0} & $\frac{-1}{3} $ & {\small J}$_{%
\text{S,}\Gamma }\text{ =2}$ & {\small 0} & {\small 0} & $\text{d}_{\Omega }%
\text{(}${\small 3193}$\text{+}\Delta \text{)}$ \\ \hline
$\text{E}_{N}\text{=}\frac{\text{25}}{2}$ & $\text{4813}$ & $(${\small 3,3,0}%
$)$ & {\small +1} & {\small 0} & {\small -1} & $\frac{-1}{3}$ & {\small J}$_{%
\text{S,N}}\text{ =3}$ & {\small 0} & {\small 100} & $\text{d}_{b}\text{(}$%
{\small 4913+}$\Delta \text{)}$ \\ \hline
$\text{E}_{\Gamma }\text{=18}$ & $\text{6793}$ & $(${\small -3,-3,0}$)$ & 
{\small -1} & {\small -3} & {\small 0} & $\frac{-1}{3}$ & {\small J}$_{\text{%
S,}\Gamma }\text{ =3}$ & {\small 0} & {\small -300} & $\text{d}_{\Omega }%
\text{(}${\small 6493}$\text{+}\Delta \text{)}$ \\ \hline
$\text{E}_{N}\text{=}\frac{\text{49}}{2}$ & $\text{9133}$ & $(${\small 4,4,0}%
$)$ & {\small +1} & {\small 0} & {\small -1} & $\frac{-1}{3}$ & {\small J}$_{%
\text{S,N}}\text{ =4}$ & {\small 0} & {\small 200} & $\text{d}_{b}\text{(%
{\small 9333}+}\Delta \text{)}$ \\ \hline
\end{tabular}
$%
\end{tabular}

As with Table 2 and Table 3, we can deduce the lowest energy quarks of all
kinds shown in Table 4A and Table 4B.

\begin{tabular}{l}
\ \ \ \ Table 4A The Lowest Energy (Ground) Quarks of all Kinds \\ 
$
\begin{tabular}{|l|l|l|l|l|l|l|l|l|l|}
\hline
$q_{\text{Name}}^{\text{I}_{\text{Z}}^{^{\prime }}}$ & q$_{N}^{\frac{1}{2}}$
& q$_{N}^{\frac{-1}{2}}$ & q$_{\Delta }^{\frac{3}{2}}$ & q$_{\Delta }^{\frac{%
1}{2}}$ & q$_{\Delta }^{\frac{-1}{2}}$ & q$_{\Delta }^{\frac{-3}{2}}$ & q$%
_{\Sigma }^{1}$ & \ q$_{\Sigma }^{0}$ & q$_{\Sigma }^{-1}$ \\ \hline
S & 0 & 0 & 0 & 0 & 0 & 0 & -1 & -1 & -1 \\ \hline
C & 0 & 0 & 0 & 0 & 0 & 0 & 0 & 0 & 0 \\ \hline
b & 0 & 0 & 0 & 0 & 0 & 0 & 0 & 0 & 0 \\ \hline
I & $\frac{1}{2}$ & $\frac{1}{2}$ & $\frac{3}{2}$ & $\frac{3}{2}$ & $\frac{3%
}{2}$ & $\frac{3}{2}$ & 1 & 1 & 1 \\ \hline
I$_{z}${\small [}{\tiny Q-}$\frac{1}{2}${\tiny (B+S}$_{G}${\tiny )}{\small ]}%
$^{\#}$ & $\frac{1}{2}$ & -$\frac{1}{2}$ & $\frac{1}{2}$ & $\frac{1}{2}$ & -$%
\frac{1}{2}$ & -$\frac{1}{2}$ & 1 & 0 & 0 \\ \hline
$\text{Q}_{q}$ & $\frac{2}{3}$ & -$\frac{1}{3}$ & $\frac{2}{3}$ & $\frac{2}{3%
}$ & $\frac{-1}{3}$ & $\frac{-1}{3}$ & $\frac{2}{3}$ & $\frac{-1}{3}$ & $%
\frac{-1}{3}$ \\ \hline
m (Mev)$^{\$}$ & 313 & 313 & 673 & 673 & 673 & 673 & 583 & 583 & 583 \\ 
\hline
$\epsilon _{\text{Name}}^{\text{I}_{\text{Z}}}$ & u$^{\frac{1}{2}}$ & d$^{%
\frac{-1}{2}}$ & u$_{\Delta }^{\frac{1}{2}}$ & u$_{\Delta }^{\frac{1}{2}}$ & 
d$_{\Delta }^{\frac{-1}{2}}$ & d$_{\Delta }^{\frac{-1}{2}}$ & u$_{\Sigma
}^{1}$ & d$_{\Sigma }^{0}$ & d$_{\Sigma }^{0}$ \\ \hline
\end{tabular}
$ \\ 
I$_{Z}^{^{\prime }}${\small = I, I-1, I-2, ... -I}. $^{\#}${\small I}$_{Z}$%
{\small \ = Q}$_{q}${\small -}$\frac{1}{2}${\small (B+S+C+b).\ }$^{\$}$%
{\small m}$^{\ast }${\small = m + }$\Delta \ \rightarrow ${\small \ m = m}$%
^{\ast }${\small - }$\Delta ${\small . \ \ \ }
\end{tabular}

\begin{tabular}{l}
\ \ \ \ \ \ \ \ Table 4B. The Lowest Energy (Ground) Quarks of all Kinds \\ 
\begin{tabular}{|l|l|l|l|l|l|l|l|l|l|}
\hline
$q_{\text{Name}}^{\text{I}_{Z}^{^{\prime }}}$ & q$_{\Xi }^{\frac{3}{2}}$ & q$%
_{\Xi }^{\frac{1}{2}}$ & q$_{S}^{0}$ & q$_{\Omega }^{0}$ & q$_{C}^{0}$ & q$%
_{b}^{0}$ & q$_{\Omega _{C}}^{0}$ & q$_{\Xi _{C}}^{\frac{1}{2}}$ & q$_{\Xi
_{C}}^{\frac{-1}{2}}$ \\ \hline
S & -2 & -2 & -1 & -3 & 0 & 0 & -2 & -1 & -1 \\ \hline
C & 0 & 0 & 0 & 0 & 1 & 0 & 1 & 1 & 1 \\ \hline
b & 0 & 0 & 0 & 0 & 0 & -1 & 0 & 0 & 0 \\ \hline
I & $\frac{1}{2}$ & $\frac{1}{2}$ & 0 & 0 & 0 & 0 & 0 & $\frac{1}{2}$ & $%
\frac{1}{2}$ \\ \hline
I$_{z}${\small [}{\tiny Q-}$\frac{1}{2}${\tiny (B+S}$_{G}${\tiny )}{\small ]}%
$^{\#}$ & $\frac{3}{2}$ & $\frac{1}{2}$ & 0 & 1 & 0 & 0 & 0 & $\frac{1}{2}$
& -$\frac{1}{2}$ \\ \hline
$\text{Q}_{q}$ & $\frac{2}{3}$ & -$\frac{1}{3}$ & -$\frac{1}{3}$ & -$\frac{1%
}{3}$ & $\frac{2}{3}$ & -$\frac{1}{3}$ & -$\frac{1}{3}$ & $\frac{2}{3}$ & -$%
\frac{1}{3}$ \\ \hline
m (Mev)$^{\$}$ & 673 & 673 & 493 & 1033 & 1753 & 4913 & 2133 & 1873 & 1873
\\ \hline
$\epsilon _{\text{Name}}^{\text{I}_{\text{Z}}}$ & u$_{\Xi }^{\frac{3}{2}}$ & 
d$_{\Xi }^{\frac{1}{2}}$ & d$_{S}^{0}$ & d$_{\Omega }^{1}$ & u$_{C}^{0}$ & d$%
_{b}^{0}$ & d$_{\Omega _{C}}^{0}$ & u$_{\Xi _{C}}^{\frac{1}{2}}$ & d$_{\Xi
_{C}}^{\frac{-1}{2}}$ \\ \hline
\end{tabular}
\\ 
$\text{I}_{Z}^{^{\prime }}${\small = I, I-1, , I-2,... -I}. $\ ^{\#}${\small %
I}$_{Z}${\small \ = Q}$_{q}${\small -}$\frac{1}{2}${\small (B+S+C+b). \ \ }$%
^{\$}${\small \ m}$^{\ast }${\small = m + }$\Delta \ \rightarrow ${\small \
m = m}$^{\ast }${\small - }$\Delta ${\small . \ \ \ }
\end{tabular}

\section{Deducing the Ground Baryons of all Kinds}

According to the Quark Model \cite{Quark Model}, a baryon is composed of
three quarks with different colors. For each flavor, the three different
colored quarks have the same I, S, C, b, Q and m. Thus we can omit the color
when we deduce the rest masses and intrinsic quantum numbers of the baryons.
We must remember, however, that three colored quarks compose a colorless
baryon. For the lowest energy baryons, the sum laws are:

\begin{equation}
\begin{tabular}{l}
$\text{S}_{\text{B}}\text{ }\text{= S}_{q_{1}}\text{+ S}_{q_{_{N}(313)}}%
\text{+ S}_{q_{_{N}(313)}}=\text{ S}_{q_{1(m)}}\text{,}$ \\ 
$\text{C}_{\text{B\ }}\text{ }\text{= C}_{q_{1}}\text{ + C}_{q_{_{N}(313)}}%
\text{ + C}_{q_{_{N}(313)}}\text{= C}_{q_{1}(m)}\text{,}$ \\ 
$\text{b}_{\text{B}}\text{=}\text{ b}_{q_{1}}\text{+ b}_{q_{_{N}(313)}}\text{%
+ b}_{q_{_{N}(313)}}\text{= b}_{q_{1(m)}}\text{,}$ \\ 
$\text{Q}_{\text{B}}\text{ = Q}_{q_{1}}\text{+ Q}_{q_{_{N}(313)}}\text{+ Q}%
_{_{q_{_{N}(313)}}}.$ \\ 
M$_{\text{B}}$ = \ $\text{m}_{q_{1}}\text{+ m}_{q_{_{N}(313)}}\text{+ m}%
_{_{q_{_{N}(313)}}}$%
\end{tabular}
\label{Sum(SCbQM)}
\end{equation}
where S$_{\text{B}}$ is the baryon's strange number, C$_{\text{B}}$ is the
baryon's charmed number, b$_{\text{B}}$ is the baryon's bottom number, Q$_{%
\text{B}}$ is the baryon's electric charge and M$_{\text{B}}$ is the
baryon's rest mass. There are strong interactions among the three quarks
(colors), but we do not know how strong. Since the rest masses of the quarks
in a baryon are large (from $\Delta $) and the rest mass of the baryon
composed by three quarks is not, we infer that there will be a strong
binding energy (E$_{Bind}$ = - 3$\Delta $) to cancel 3$\Delta $ from the
three quarks: M$_{\text{B}}\text{ = m}_{q_{1}}^{\ast }$ + \ m$%
_{q_{_{N}(313)}}^{\ast }$ \ + m$_{_{q_{_{N}(313)}}}^{\ast }$- $\left\vert 
\text{E}_{Bind}\right\vert $ = m$_{q_{1}}$+ m$_{q_{_{N}(313)}}$+ m$%
_{_{q_{_{N}(313)}}}$+ 3$\Delta $ - 3$\Delta $ = m$_{q_{1}}$+ m$%
_{q_{_{N}(313)}}$+ m$_{_{q_{_{N}(313)}}}$.

Using (\ref{Sum(SCbQM)}) and Table 4(A and B), we can find the rest masses
and the intrinsic\ quantum numbers (I, S, C, b and Q) of the lowest energy
(ground) baryons of all kinds shown in Table 5. The experimental results are
from \cite{Baryon04}

\begin{tabular}{l}
\ \ \ \ \ \ \ \ \ \ \ Table 5. The Ground Baryons (Lowest Energy) of all
Kinds \\ 
\begin{tabular}{|l|l|l|l|l|l|l|l|l|l|l|l|}
\hline
{\small q}$_{i}^{I_{z}}$ & {\small q}$_{j}$ & {\small q}$_{k}$ & {\small I}
& {\small S} & {\small C} & {\small b} & {\small Q} & {\small M} & {\small %
Baryon} & {\small Exper.} & $\frac{\Delta M}{M}${\small \%} \\ \hline
{\small u}$^{\frac{1}{2}}${\small (313)} & {\small u} & {\small d} & $\frac{1%
}{2}$ & {\small 0} & {\small 0} & {\small 0} & {\small 1} & {\small 939} & 
{\small p(939)} & {\small p(938)} & {\small 0.11} \\ \hline
{\small d}$^{\frac{-1}{2}}${\small (313)} & {\small u} & {\small d} & $\frac{%
1}{2}$ & {\small 0} & {\small 0} & {\small 0} & {\small 0} & {\small 939} & 
{\small n(939)} & {\small n(940)} & {\small 0.11} \\ \hline
{\small d}$_{s}^{0}${\small (493)} & {\small u} & {\small d} & {\small 0} & 
{\small -1} & {\small 0} & {\small 0} & {\small 0} & {\small 1119} & $%
\Lambda ${\small (1119)} & $\Lambda ^{0}(1116)$ & {\small 0.27} \\ \hline
{\small u}$_{c}^{0}${\small (1753)} & {\small u} & {\small d} & {\small 0} & 
{\small 0} & {\small 1} & {\small 0} & {\small 1} & {\small 2379} & $\Lambda
_{c}${\small (2379)} & $\Lambda _{c}^{+}(2285) $ & {\small 4.1} \\ \hline
{\small u}$_{c}^{0}${\small (1753)} & {\small u} & {\small u} & {\small 1} & 
{\small 0} & {\small 1} & {\small 0} & {\small 2} & {\small 2379} & $\Sigma
_{c}^{++}${\small (2379)} & $\Sigma _{c}^{++}${\small (2455)} & {\small 3.1}
\\ \hline
{\small u}$_{c}^{0}${\small (1753)} & {\small u} & {\small d} & {\small 1} & 
{\small 0} & {\small 1} & {\small 0} & {\small 1} & {\small 2379} & $\Sigma
_{c}^{+}${\small (2379)} & $\Sigma _{c}^{+}${\small (2455)} & {\small 3.1}
\\ \hline
{\small u}$_{c}^{0}${\small (1753)} & {\small d} & {\small d} & {\small 1} & 
{\small 0} & {\small 1} & {\small 0} & {\small 0} & {\small 2379} & $\Sigma
_{c}^{-}${\small (2379)} & $\Sigma _{c}^{-}${\small (2455)} & {\small 3.1}
\\ \hline
{\small d}$_{b}^{0}${\small (4913)} & {\small u} & {\small d} & {\small 0} & 
{\small 0} & {\small 0} & {\small -1} & {\small 0} & {\small 5539} & $%
\Lambda _{b}${\small (2379)} & $\Lambda _{b}^{0}(5624)$ & {\small 1.5} \\ 
\hline
{\small u}$_{\Sigma }^{1}${\small (583)} & {\small u} & {\small d} & {\small %
1} & {\small -1} & {\small 0} & {\small 0} & {\small 1} & {\small 1209} & $%
\Sigma ^{+}${\small (1209)} & $\Sigma ^{+}(1189)$ & {\small 1.7} \\ \hline
{\small d}$_{\Sigma }^{0}${\small (583)} & {\small u} & {\small d} & {\small %
1} & {\small -1} & {\small 0} & {\small 0} & {\small 0} & {\small 1209} & $%
\Sigma ^{0}${\small (1209)} & $\Sigma ^{0}(1193)$ & {\small 1.4} \\ \hline
{\small d}$_{\Sigma }^{0}${\small (583)} & {\small d} & {\small d} & {\small %
1} & {\small -1} & {\small 0} & {\small 0} & {\small -1} & {\small 1209} & $%
\Sigma ^{-1`}${\small (1209)} & $\Sigma ^{-}(1197)$ & {\small 1.0} \\ \hline
{\small u}$_{\Xi }^{\frac{3}{2}}${\small (673)} & {\small d} & {\small d} & $%
\frac{1}{2}$ & {\small -2} & {\small 0} & {\small 0} & {\small 0} & {\small %
1299} & $\Xi ^{0}${\small (1299)} & $\Xi ^{0}(1315)$ & {\small 1.2} \\ \hline
{\small d}$_{\Xi }^{\frac{1}{2}}${\small (673)} & {\small d} & {\small d} & $%
\frac{1}{2}$ & {\small -2} & {\small 0} & {\small 0} & {\small -1} & {\small %
1299} & $\Xi ^{-}${\small (1299)} & $\Xi ^{-}(1321)$ & {\small 1.7} \\ \hline
{\small d}$_{\Omega }^{1}${\small (1033)} & {\small d} & {\small d} & 
{\small 0} & {\small -3} & {\small 0} & {\small 0} & {\small -1} & {\small %
1659} & $\Omega ^{-}${\small (1659)} & $\Omega ^{-}(1672)$ & {\small 0.78}
\\ \hline
{\small d}$_{\Omega _{c}}^{0}${\small (2133)} & {\small u} & {\small d} & 
{\small 0} & {\small -2} & {\small 1} & {\small 0} & {\small 0} & {\small %
2759} & $\Omega _{C}${\small (2759)} & $\Omega _{C}${\small (2698)} & 
{\small 2.0} \\ \hline
{\small u}$_{\Xi _{C}}^{\frac{1}{2}}${\small (1873)} & {\small u} & {\small d%
} & $\frac{1}{2}$ & {\small -1} & {\small 1} & {\small 0} & {\small 1} & 
{\small 2499} & $\Xi _{C}${\small (2499)} & $\Xi _{c}^{+}(2466)$ & {\small %
1.4} \\ \hline
{\small d}$_{\Xi _{C}}^{\frac{-1}{2}}${\small (1873)} & {\small u} & {\small %
d} & $\frac{1}{2}$ & {\small -1} & {\small 1} & {\small 0} & {\small 0} & 
{\small 2499} & $\Xi _{C}${\small (2499)} & $\Xi _{c}^{0}(2472)$ & {\small %
1.2.} \\ \hline
{\small u}$_{\Delta }^{\frac{1}{2}}${\small (673)} & {\small u} & {\small u}
& $\frac{3}{2}$ & {\small 0} & {\small 0} & {\small 0} & {\small \ 2} & 
{\small 1299} & $\Delta ^{++}(1299)$ & $\Delta ^{++}(1232)$ & $\Gamma $%
{\small =120} \\ \hline
{\small u}$_{\Delta }^{\frac{1}{2}}${\small (673)} & {\small u} & {\small d}
& $\frac{3}{2}$ & {\small 0} & {\small 0} & {\small 0} & {\small \ 1} & 
{\small 1299} & $\Delta ^{+}(1299)$ & $\Delta ^{+}(1232)$ & $\Gamma ${\small %
=120} \\ \hline
{\small d}$_{\Delta }^{\frac{-1}{2}}${\small (673)} & {\small u} & {\small d}
& $\frac{3}{2}$ & {\small 0} & {\small 0} & {\small 0} & {\small \ 0} & 
{\small 1299} & $\Delta ^{0}(1299)$ & $\Delta ^{0}(1232)$ & $\Gamma ${\small %
=120} \\ \hline
{\small d}$_{\Delta }^{\frac{-1}{2}}${\small (673)} & {\small d} & {\small d}
& $\frac{3}{2}$ & {\small 0} & {\small 0} & {\small 0} & {\small -1} & 
{\small 1299} & $\Delta ^{-}(1299)$ & $\Delta ^{-}(1232)$ & $\Gamma ${\small %
=120} \\ \hline
\end{tabular}
\\ 
\ \ In the Table, u $\equiv $ {\small u}$^{\frac{1}{2}}${\small (313) and d} 
$\equiv $ {\small d}$^{\frac{-1}{2}}${\small (313).}
\end{tabular}

Table 5 shows that the intrinsic quantum numbers (I, S, C, b and Q) of the
deduced baryons are the same as the experimental results and the deduced
masses are consistent with the experimental results.

\section{Predictions\ \ \ \ \ \ \ \ \ \ \ \ \ \ \ \ \ \ \ \ \ \ \ \ \ \ \ \
\ \ \ \ \ \ \ \ \ \ \ \ \ \ \ \ \ \ \ \ \ \ \ \ \ \ \ \ \ \ \ \ \ \ \ \ \ \
\ \ \ \ \ \ \ \ \ \ \ \ \ \ \ \ \ \ \ \ \ \ \ \ \ \ \ \ \ \ \ \ \ \ \ \ \ \
\ \ \ \ \ \ \ \ \ \ \ \ \ \ \ \ \ \ \ \ \ \ \ \ \ \ \ \ \ \ \ \ \ \ \ \ \ \
\ \ \ \ \ \ \ \ \ \ \ \ \ \ \ \ \ \ \ \ \ \ \ \ \ \ \ \ \ \ \ \ \ \ \ \ \ \
\ \ \ \ \ \ \ \ \ \ \ \ \ \ \ \ \ \ \ \ \ \ \ \ \ \ \ \ \ \ \ \ \ \ \ \ \ \
\ \ \ \ \ \ \ \ \ \ \ \ \ \ \ \ \ \ \ \ \ \ \ \ \ \ \ \ \ \ \ \ \ \ \ \ \ \
\ \ \ \ \ \ \ \ \ \ \ \ \ \ \ \ \ \ \ \ \ \ \ \ \ \ \ \ \ \ \ \ \ \ \ \ \ \
\ \ \ \ \ \ \ \ \ \ \ \ \ \ \ \ \ \ \ \ \ \ \ \ \ \ \ \ \ \ \ \ \ \ \ \ \ \
\ \ \ \ \ \ \ \ \ \ \ \ \ \ \ \ \ \ \ \ \ \ \ \ \ \ \ \ \ \ \ \ \ \ \ \ \ \
\ \ \ \ \ \ \ \ \ \ \ \ \ \ \ \ \ \ \ \ \ \ \ \ \ \ \ \ \ \ \ \ \ \ \ \ \ \
\ \ \ \ \ \ \ \ \ \ \ \ \ \ \ \ \ \ \ \ \ \ \ \ \ \ \ \ \ \ \ \ \ \ \ \ \ \
\ \ \ \ \ \ \ \ \ \ \ \ \ \ \ \ \ \ \ \ \ \ \ \ \ \ \ \ \ \ \ \ \ \ \ \ \ \
\ \ \ \ \ \ \ \ \ \ \ \ \ \ \ \ \ \ \ \ \ \ \ \ \ \ \ \ \ \ \ \ \ \ \ \ \ \
\ \ \ \ \ \ \ \ \ \ \ \ \ \ \ \ \ \ \ \ \ \ \ \ \ \ \ \ \ \ \ \ \ \ \ \ \ \
\ \ \ \ \ \ \ \ \ \ \ \ \ \ \ \ \ \ \ \ \ \ \ \ \ \ \ \ \ \ \ \ \ \ \ \ \ \
\ \ \ \ \ \ \ \ \ \ \ \ \ \ \ \ \ \ \ \ \ \ \ \ \ \ \ \ \ \ \ \ \ \ \ \ \ \
\ \ \ \ \ \ \ \ \ \ \ \ \ \ \ \ \ \ \ \ \ \ \ \ \ \ \ \ \ \ \ \ \ \ \ \ \ \
\ \ \ \ \ \ \ \ \ \ \ \ \ \ \ \ \ \ \ \ \ \ \ \ \ \ \ \ \ \ \ \ \ \ \ \ \ \
\ \ \ \ \ \ \ \ \ \ \ \ \ \ \ \ \ \ \ \ \ \ \ \ \ \ \ \ \ \ \ \ \ \ \ \ \ \
\ \ \ \ \ \ \ \ \ \ \ \ \ \ \ \ \ \ \ \ \ \ \ \ \ \ \ \ \ \ \ \ \ \ \ \ \ \
\ \ \ \ \ \ \ \ \ \ \ \ \ \ \ \ \ \ \ \ \ \ \ \ \ \ \ \ \ \ \ \ \ \ \ \ \ \
\ \ \ \ \ \ \ \ \ \ \ \ \ \ \ \ \ \ \ \ \ \ \ \ \ \ \ \ \ \ \ \ \ \ \ \ \ \
\ \ \ \ \ \ \ \ \ \ \ \ \ \ \ \ \ \ \ \ \ \ \ \ \ \ \ \ \ \ \ \ \ \ \ \ \ \
\ \ \ \ \ \ \ \ \ \ \ \ \ \ \ \ \ \ \ \ \ \ \ \ \ \ \ \ \ \ \ \ \ \ \ \ \ \
\ \ \ \ \ \ \ \ \ \ \ \ \ \ \ \ \ \ \ \ \ \ \ \ \ \ \ \ \ \ \ \ \ \ \ \ \ \
\ \ \ \ \ \ \ \ \ \ \ \ \ \ \ \ \ \ \ \ \ \ \ \ \ \ \ \ \ \ \ \ \ \ \ \ \ \
\ \ \ \ \ \ \ \ \ \ \ \ \ \ \ \ \ \ \ \ \ \ \ \ \ \ \ \ \ \ \ \ \ \ \ \ \ \
\ \ \ \ \ \ \ \ \ \ \ \ \ \ \ \ \ \ \ \ \ \ \ \ \ \ \ \ \ \ \ \ \ \ \ \ \ \
\ \ \ \ \ \ \ \ \ \ \ \ \ \ \ \ \ \ \ \ \ \ \ \ \ \ \ \ \ \ \ \ \ \ \ \ \ \
\ \ \ \ \ \ \ \ \ \ \ \ \ \ \ \ \ \ \ \ \ \ \ \ \ \ \ \ \ \ \ \ \ \ \ \ \ \
\ \ \ \ \ \ \ \ \ \ \ \ \ \ \ \ \ \ \ \ \ \ \ \ \ \ \ \ \ \ \ \ \ \ \ \ \ \
\ \ \ \ \ \ \ \ \ \ \ \ \ \ \ \ \ \ \ \ \ \ \ \ \ \ \ \ \ \ \ \ \ \ \ \ \ \
\ \ \ \ \ \ \ \ \ \ \ \ \ \ \ \ \ \ \ \ \ \ \ \ \ \ \ \ \ \ \ \ \ \ \ \ \ \
\ \ \ \ \ \ \ \ \ \ \ \ \ \ \ \ \ \ \ \ \ \ \ \ \ \ \ \ \ \ \ \ \ \ \ \ \ \
\ \ \ \ \ \ \ \ \ \ \ \ \ \ \ \ \ \ \ \ \ \ \ \ \ \ \ \ \ \ \ \ \ \ \ \ \ \
\ \ \ \ \ \ \ \ \ \ \ \ \ \ \ \ \ \ \ \ \ \ \ \ \ \ \ \ \ \ \ \ \ \ \ \ \ \
\ \ \ \ \ \ \ \ \ \ \ \ \ \ \ \ \ \ \ \ \ \ \ \ \ \ \ \ \ \ \ \ \ \ \ \ \ \
\ \ \ \ \ \ \ \ \ \ \ \ \ \ \ \ \ \ \ \ \ \ \ \ \ \ \ \ \ \ \ \ \ \ \ \ \ \
\ \ \ \ \ \ \ \ \ \ \ \ \ \ \ \ \ \ \ \ \ \ \ \ \ \ \ \ \ \ \ \ \ \ \ \ \ \
\ \ \ \ \ \ \ \ \ \ \ \ \ \ \ \ \ \ \ \ \ \ \ \ \ \ \ \ \ \ \ \ \ \ \ \ \ \
\ \ \ \ \ \ \ \ \ \ \ \ \ \ \ \ \ \ \ \ \ \ \ \ \ \ \ \ \ \ \ \ \ \ \ \ \ \
\ \ \ \ \ \ \ \ \ \ \ \ \ \ \ \ \ \ \ \ \ \ \ \ \ \ \ \ \ \ \ \ \ \ \ \ \ \
\ \ \ \ \ \ \ \ \ \ \ \ \ \ \ \ \ \ \ \ \ \ \ \ \ \ \ \ \ \ \ \ \ \ \ \ \ \
\ \ \ \ \ \ \ \ \ \ \ \ \ \ \ \ \ \ \ \ \ \ \ \ \ \ \ \ \ \ \ \ \ \ \ \ \ \
\ \ \ \ \ \ \ \ \ \ \ \ \ \ \ \ \ \ \ \ \ \ \ \ \ \ \ \ \ \ \ \ \ \ \ \ \ \
\ \ \ \ \ \ \ \ \ \ \ \ \ \ \ \ \ \ \ \ \ \ \ \ \ \ \ \ \ \ \ \ \ \ \ \ \ \
\ \ \ \ \ \ \ \ \ \ \ \ \ \ \ \ \ \ \ \ \ \ \ \ \ \ \ \ \ \ \ \ \ \ \ \ \ \
\ \ \ \ \ \ \ \ \ \ \ \ \ \ \ \ \ \ \ \ \ \ \ \ \ \ \ \ \ \ \ \ \ \ \ \ \ \
\ \ \ \ \ \ \ \ \ \ \ \ \ \ \ \ \ \ \ \ \ \ \ \ \ \ \ \ \ \ \ \ \ \ \ \ \ \
\ \ \ \ \ \ \ \ \ \ \ \ \ \ \ \ \ \ \ \ \ \ \ \ \ \ \ \ \ \ \ \ \ \ \ \ \ \
\ \ \ \ \ \ \ \ \ \ \ \ \ \ \ \ \ \ \ \ \ \ \ \ \ \ \ \ \ \ \ \ \ \ \ \ \ \
\ \ \ \ \ \ \ \ \ \ \ \ \ \ \ \ \ \ \ \ \ \ \ \ \ \ \ \ \ \ \ \ \ \ \ \ \ \
\ \ \ \ \ \ \ \ \ \ \ \ \ \ \ \ \ \ \ \ \ \ \ \ \ \ \ \ \ \ \ \ \ \ \ \ \ \
\ \ \ \ \ \ \ \ \ \ \ \ \ \ \ \ \ \ \ \ \ \ \ \ \ \ \ \ \ \ \ \ \ \ \ \ \ \
\ \ \ \ \ \ \ \ \ \ \ \ \ \ \ \ \ \ \ \ \ \ \ \ \ \ \ \ \ \ \ \ \ \ \ \ \ \
\ \ \ \ \ \ \ \ \ \ \ \ \ \ \ \ \ \ \ \ \ \ \ \ \ \ \ \ \ \ \ \ \ \ \ \ \ \
\ \ \ \ \ \ \ \ \ \ \ \ \ \ \ \ \ \ \ \ \ \ \ \ \ \ \ \ \ \ \ \ \ \ \ \ \ \
\ \ \ \ \ \ \ \ \ \ \ \ \ \ \ \ \ \ \ \ \ \ \ \ \ \ \ \ \ \ \ \ \ \ \ \ \ \
\ \ \ \ \ \ \ \ \ \ \ \ \ \ \ \ \ \ \ \ \ \ \ \ \ \ \ \ \ \ \ \ \ \ \ \ \ \
\ \ \ \ \ \ \ \ \ \ \ \ \ \ \ \ \ \ \ \ \ \ \ \ \ \ \ \ \ \ \ \ \ \ \ \ \ \
\ \ \ \ \ \ \ \ \ \ \ \ \ \ \ \ \ \ \ \ \ \ \ \ \ \ \ \ \ \ \ \ \ \ \ \ \ \
\ \ \ \ \ \ \ \ \ \ \ \ \ \ \ \ \ \ \ \ \ \ \ \ \ \ \ \ \ \ \ \ \ \ \ \ \ \
\ \ \ \ \ \ \ \ \ \ \ \ \ \ \ \ \ \ \ \ \ \ \ \ \ \ \ \ \ \ \ \ \ \ \ \ \ \
\ \ \ \ \ \ \ \ \ \ \ \ \ \ \ \ \ \ \ \ \ \ \ \ \ \ \ \ \ \ \ \ \ \ \ \ \ \
\ \ \ \ \ \ \ \ \ \ \ \ \ \ \ \ \ \ \ \ \ \ \ \ \ \ \ \ \ \ \ \ \ \ \ \ \ \
\ \ \ \ \ \ \ \ \ \ \ \ \ \ \ \ \ \ \ \ \ \ \ \ \ \ \ \ \ \ \ \ \ \ \ \ \ \
\ \ \ \ \ \ \ \ \ \ \ \ \ \ \ \ \ \ \ \ \ \ \ \ \ \ \ \ \ \ \ \ \ \ \ \ \ \
\ \ \ \ \ \ \ \ \ \ \ \ \ \ \ \ \ \ \ \ \ \ \ \ \ \ \ \ \ \ \ \ \ \ \ \ \ \
\ \ \ \ \ \ \ \ \ \ \ \ \ \ \ \ \ \ \ \ \ \ \ \ \ \ \ \ \ \ \ \ \ \ \ \ \ \
\ \ \ \ \ \ \ \ \ \ \ \ \ \ \ \ \ \ \ \ }

From Table 2, Table 3, Table 4A and Table 4B, we can predict some quarks and
baryons as shown in the following list:

\begin{tabular}{|l|l|l|l|l|l|l|}
\hline
{\small q}$_{i}$(Quark$^{\#}$) & $\text{d}_{S}\text{(773) }$ & $\text{d}_{S}%
\text{(3753)}$ & $\text{d}_{S}\text{(9613)}$ & $\text{u}_{C}\text{(6073)}$ & 
$\text{d}_{b}\text{(9333)}$ & $\text{d}_{\Omega }\text{(3193)}$ \\ \hline
{\small q}$_{j}$ & u(313) & u(313) & u(313) & u(313) & u(313) & d(313) \\ 
\hline
{\small q}$_{k}$ & d(313) & d(313) & d(313) & d(313) & d(313) & d(313) \\ 
\hline
Prediction & $\Lambda ^{0}$(1399) & $\Lambda ^{0}$(4379) & $\Lambda ^{0}$%
(10239) & $\Lambda _{c}^{+}$(6699) & $\Lambda _{b}^{0}$(9959) & $\Omega ^{-}$%
(3819) \\ \hline
S, C, b & $\Lambda ^{0}$(1406)$^{\$}$ & S = -1 & S = -1 & C = +1 & b = 1 & S
=-3 \\ \hline
\end{tabular}

\ $^{\#}\ $Predicted\ quarks$\ \ \ \ ^{\$}\Lambda ^{0}$(1399) has
discovered-- $\Lambda ^{0}$(1406) with ($\frac{\Delta M}{M}$\%) = 0.5\%.

\section{Discussion\ \ \ \ \ \ \ \ \ \ \ \ \ \ \ \ \ \ \ \ \ \ \ \ \ \ \ \ \
\ \ \ \ }

1. The fact that physicists have not found any free quark shows that the
binding energies are strong. The binding energy (-3$\Delta $) is a
phenomenological approximation of the color's strong interaction energy. The
binding energy (-3$\Delta $) is always cancelled by the corresponding parts
(3$\Delta $) of the rest masses of the three quarks. Thus we can omit the
binding energy and the corresponding rest mass parts of the three quarks.
This effect makes it appear as if there is no binding energy in baryons.

2. The energy band excited quarks of the elementary quark $\epsilon _{u}$(or 
$\epsilon _{d}$) are the short-lived and scarce quarks, such as d$_{s}$%
(493), u$_{c}$(1753), d$_{b}$(4913), q$_{\Xi }$(673), q$_{\Sigma }$(583) and
q$_{\Delta }$(673). The q$_{N}(313)$ corresponding to the energy band with $%
\overrightarrow{n}$ = (0, 0, 0) will be a short-lived and scarce quark. It
is, however, a lowest energy quark with strong binding energy. Since there
is no lower energy position that they can decay to, they are not short-lived
quarks. Because they have the same rest mass and intrinsic quantum numbers
as the free excited quarks u(313) and d(313), they cannot be distinguished
from u(313) and d(313) by experiments; therefore, they are not scarce
quarks. The u(313) and d(313) with $\overrightarrow{n}$ = (0, 0, 0) will be
covered up by free excited u(313) and d(313) in experiments. Thus, we can
omit u(313) and d(313) with $\overrightarrow{n}$ = (0, 0, 0). There are only
long-lived free excited the u(313)-quark and the d(313)-quark in both theory
and experiments. \ \ \ \ \ \ \ \ \ \ \ \ \ \ \ \ \ \ \ \ \ \ \ \ \ \ \ \ \ \
\ \ \ \ \ \ \ \ \ \ \ \ \ \ \ \ \ \ \ \ \ \ \ \ \ \ \ \ \ \ \ \ \ \ \ \ \ \
\ \ \ \ \ \ \ \ \ \ \ \ \ \ \ \ \ \ \ \ \ \ \ \ \ \ \ \ \ \ \ \ \ \ \ \ \ \
\ \ \ \ \ \ \ \ \ \ \ \ \ \ \ \ \ \ \ \ \ \ \ \ \ \ \ \ \ \ \ \ \ \ \ \ \ \
\ \ \ \ \ \ \ \ \ \ \ \ \ \ \ \ \ \ \ \ \ \ \ \ \ \ \ \ \ \ \ \ \ \ \ \ \ \
\ \ \ \ \ \ \ \ \ \ \ \ \ \ \ \ \ \ \ \ \ \ \ \ \ \ \ \ \ \ \ \ \ \ \ \ \ \
\ \ \ \ \ \ \ \ \ \ \ \ \ \ \ \ \ \ \ \ \ \ \ \ \ \ \ \ \ \ \ \ \ \ \ \ \ \
\ \ \ \ \ \ \ \ \ \ \ \ \ \ \ \ \ \ \ \ \ \ \ \ \ \ \ \ \ \ \ \ \ \ \ \ \ \
\ \ \ \ \ \ \ \ \ \ \ \ \ \ \ \ \ \ \ \ \ \ \ \ \ \ \ \ \ \ \ \ \ \ \ \ \ \
\ \ \ \ \ \ \ \ \ \ \ \ \ \ \ \ \ \ \ \ \ \ \ \ \ \ \ \ \ \ \ \ \ \ \ \ \ \
\ \ \ \ \ \ \ \ \ \ \ \ \ \ \ \ \ \ \ \ \ \ \ \ \ \ \ \ \ \ \ \ \ \ \ \ \ \
\ \ \ \ \ \ \ \ \ \ \ \ \ \ \ \ \ \ \ \ \ \ \ \ \ \ \ \ \ \ \ \ \ \ \ \ \ \
\ \ \ \ \ \ \ \ \ \ \ \ \ \ \ \ \ \ \ \ \ \ \ \ \ \ \ \ \ \ \ \ \ \ \ \ \ \
\ \ \ \ \ \ \ \ \ \ \ \ \ \ \ \ \ \ \ \ \ \ \ \ \ \ \ \ \ \ \ \ \ \ \ \ \ \
\ \ \ \ \ \ \ \ \ \ \ \ \ \ \ \ \ \ \ \ \ \ \ \ \ \ \ \ \ \ \ \ \ \ \ \ \ \
\ \ \ \ \ \ \ \ \ \ \ \ \ \ \ \ \ \ \ \ \ \ \ \ \ \ \ \ \ \ \ \ \ \ \ \ \ \
\ \ \ \ \ \ \ \ \ \ \ \ \ \ \ \ \ \ \ \ \ \ \ \ \ \ \ \ \ \ \ \ \ \ \ \ \ \
\ \ \ \ \ \ \ \ \ \ \ \ \ \ \ \ \ \ \ \ \ \ \ \ \ \ \ \ \ \ \ \ \ \ \ \ \ \
\ \ \ \ \ \ \ \ \ \ \ \ \ \ \ \ \ \ \ \ \ \ \ \ \ \ \ \ \ \ \ \ \ \ \ \ \ \
\ \ \ \ \ \ \ \ \ \ \ \ \ \ \ \ \ \ \ \ \ \ \ \ \ \ \ \ \ \ \ \ \ \ \ \ \ \
\ \ \ \ \ \ \ \ \ \ \ \ \ \ \ \ \ \ \ \ \ \ \ \ \ \ \ \ \ \ \ \ \ \ \ \ \ \
\ \ \ \ \ \ \ \ \ \ \ \ \ \ \ \ \ \ \ \ \ \ \ \ \ \ \ \ \ \ \ \ \ \ \ \ \ \
\ \ \ \ \ \ \ \ \ \ \ \ \ \ \ \ \ \ \ \ \ \ \ \ \ \ \ \ \ \ \ \ \ \ \ \ \ \
\ \ \ \ \ \ \ \ \ \ \ \ \ \ \ \ \ \ \ \ \ \ \ \ \ \ \ \ \ \ \ \ \ \ \ \ \ \
\ \ \ \ \ \ \ \ \ \ \ \ \ \ \ \ \ \ \ \ \ \ \ \ \ \ \ \ \ \ \ \ \ \ \ \ \ \
\ \ \ \ \ \ \ \ \ \ \ \ \ \ \ \ \ \ \ \ \ \ \ \ \ \ \ \ \ \ \ \ \ \ \ \ \ \
\ \ \ \ \ \ \ \ \ \ \ \ \ \ \ \ \ \ \ \ \ \ \ \ \ \ \ \ \ \ \ \ \ \ \ \ \ \
\ \ \ \ \ \ \ \ \ \ \ \ \ \ \ \ \ \ \ \ \ \ \ \ \ \ \ \ \ \ \ \ \ \ \ \ \ \
\ \ \ \ \ \ \ \ \ \ \ \ \ \ \ \ \ \ \ \ \ \ \ \ \ \ \ \ \ \ \ \ \ \ \ \ \ \
\ \ \ \ \ \ \ \ \ \ \ \ \ \ \ \ \ \ \ \ \ \ \ \ \ \ \ \ \ \ \ \ \ \ \ \ \ \
\ \ \ \ \ \ \ \ \ \ \ \ \ \ \ \ \ \ \ \ \ \ \ \ \ \ \ \ \ \ \ \ \ \ \ \ \ \
\ \ \ \ \ \ \ \ \ \ \ \ \ \ \ \ \ \ \ \ \ \ \ \ \ \ \ \ \ \ \ \ \ \ \ \ \ \
\ \ \ \ \ \ 

3. The five quarks of the current Quark Model correspond to the five deduced
ground quarks [u$\leftrightarrow $u(313), d$\leftrightarrow $d(313), s$%
\leftrightarrow $d$_{s}$(493), c$\leftrightarrow $u$_{c}$(1753) and b$%
\leftrightarrow $d$_{b}$(4913)] (see Table 4B and 4B as well as Table 11 of 
\cite{0502091}). The current Quark Model uses only these five quarks to
explain baryons and mesons. Using quantized energy bands and the
phenomenological formulae, the new quark model can deduce the rest masses
and the intrinsic quantum numbers of excited quarks, baryons and mesons \cite
{0502091} from one elementary quark family. Thus, the current Quark Model is
the five ground quark approximation of the new quark model.

\section{Conclusions\ \ \ \ \ \ \ \ \ \ \ \ \ \ \ \ \ \ \ \ \ \ \ \ \ \ \ \
\ \ \ \ \ \ }

1. There is only one elementary quark family $\epsilon $ with three colors
and two isospin states ($\epsilon _{u}$ with I$_{Z}$ = $\frac{1}{2}$ and Q =
+$\frac{2}{3}$, $\epsilon _{d}$ with I$_{Z}$ = $\frac{-1}{2}$ and Q = -$%
\frac{1}{3}$) for each color. Thus there are six Fermi (s = $\frac{1}{2}$)
elementary quarks with S = C = b = 0 in the vacuum.

2. All quarks in hadrons are the excited state of the elementary quark $%
\epsilon $. There are two types of excited states: free excited states and
energy band excited states. The free excited states are only the
u(313)-quark and the d(313)-quark; the energy band excited states are the
short-lived and scarce quarks, such as d$_{s}$(493), u$_{c}$(1753), d$_{b}$%
(4913), q$_{N}$(583), q$_{\Sigma }$(583) and q$_{\Delta }$(673).

3. The short-lived and scarce quarks mainly originate from the expanded
quantization. They are the energy band excited states of the elementary
quarks.

4. There is a strong binding energy (-3$\Delta $) among the three quarks
(colors) inside a baryon. It may be a possible foundation of the quark
confinement.

5. This paper predicts new baryons $\Lambda _{c}^{+}$(6699), $\Lambda
_{b}^{0}$(9959), $\Lambda ^{0}$(4379) and $\Lambda ^{0}$(10239).

\begin{center}
\bigskip \textbf{Acknowledgments}
\end{center}

I sincerely thank Professor Robert L. Anderson for his valuable advice. I
acknowledge\textbf{\ }my indebtedness to Professor D. P. Landau for his help
also. I would like to express my heartfelt gratitude to Dr. Xin Yu for
checking the calculations. I sincerely thank Professor Yong-Shi Wu for his
important advice and help. I thank Professor Wei-Kun Ge for his support and
help. I sincerely thank Professor Kang-Jie Shi for his advice.

\bigskip

\bigskip

\end{document}